\documentclass{mem}
\usepackage{natbib}\usepackage{txfonts}\usepackage{balance}
\usepackage{graphicx}
\usepackage[a4paper,breaklinks,dvipdfm]{hyperref}
\idline{85}{0}
\begin{document}

\newcommand{\epm}{\ensuremath{e^{\pm}\;}}
\newcommand{\sigv}{\ensuremath{\langle \sigma v\rangle}}
\newcommand{\omh}{\ensuremath{\Omega_{\rm CDM} h^{2}}}
\newcommand{\omb}{\ensuremath{\Omega_{\rm B} h^{2}}}
\newcommand{\neff}{\ensuremath{{\rm N}_{\rm eff}}}
\newcommand{\Deln}{\ensuremath{\Delta{\rm N}_\nu}}
\newcommand{\nnu}{\ensuremath{N_\nu}}
\newcommand{\mchi}{\ensuremath{m_\chi}}
\newcommand{\II}{\hspace{0.5ex}{\rm{I}\hspace*{-1.25ex} \rm{I}\hspace*{1.25ex}}}
\newcommand{\beq}{\begin{equation}}  
\newcommand{\eeq}{\end{equation}}

\def\3he{$^3$He}
\def\4he{$^4$He}
\def\7li{$^7$Li}
\def\Yp{Y$_{\rm P}$}
\def\yd{$y_{\rm DP}$}
\def\hii{H\thinspace{$\scriptstyle{\rm II}$}}
\def\hi{H\thinspace{$\scriptstyle{\rm I}$}}
\newcommand{\ie}{{\it i.e.}}
\newcommand{\eg}{{\it e.g.}}
\newcommand{\etal}{{\it et al.}}

\title{
Light WIMPs, Equivalent Neutrinos, BBN, And The CMB
}

   \subtitle{}

\author{
Gary Steigman\inst{1,2,3} 
\and Kenneth M. Nollett\inst{2,4}
          }

  \offprints{G. Steigman}

\institute{
Physics Department, The Ohio State University,
Columbus, Ohio, USA
\and
Center for Cosmology and Astro-Particle Physics, 
The Ohio State University, Columbus, Ohio, USA
\and
Departamento de Astronomia, Universidade de S\~ao Paulo,
S\~ao Paulo, Brasil\\
\email{steigman.1@osu.edu}
\and
Department of Physics and Astronomy, Ohio University, Athens, Ohio, USA\\
\email{nollett@ohio.edu}
}

\authorrunning{Steigman \& Nollett}

\titlerunning{Light WIMPs}

\abstract{
Recent updates to the observational determinations of the primordial abundances of helium (\4he) and deuterium are compared to the predictions of BBN to infer the universal ratio of baryons to photons, $\eta_{10} \equiv 10^{10}(n_{\rm B}/n_{\gamma})_{0}$ (or, the present Universe baryon mass density parameter, \omb~$= \eta_{10}/273.9$) as well as to constrain the effective number of neutrinos (\neff) and the number of equivalent neutrinos (\Deln).  These BBN  results are compared to those derived independently from the Planck CMB data.  In the absence of a light WIMP (\mchi~$\ga 20\,{\rm MeV}$), \neff~$= 3.05(1 + \Deln/3)$.  In this case, there is excellent agreement between BBN and the CMB but, the joint fit reveals that \Deln~$= 0.40\pm0.17$, disfavoring standard big bang nucleosynthesis (SBBN) (\Deln~= 0) at $\sim 2.4\,\sigma$, as well as a sterile neutrino (\Deln~= 1) at $\sim 3.5\,\sigma$.  In the presence of a light WIMP (\mchi~$\la 20\,{\rm MeV}$), the relation between \neff~and \Deln~depends on the WIMP mass, leading to degeneracies among \neff, \Deln, and \mchi.  The complementary and independent BBN and CMB data can break some of these degeneracies.  Depending on the nature of the light WIMP (Majorana or Dirac fermion, real or complex scalar) the joint BBN + CMB analyses set a {\bf lower} bound to \mchi~in the range $0.5 - 5\,{\rm MeV}$ ($m_{\chi}/m_{e} \ga 1 - 10$) and, they identify {\bf best fit} values for \mchi~in the range $5 - 10\,{\rm MeV}$.  The joint BBN + CMB analyses find a {\bf best fit} value for the number of equivalent neutrinos, \Deln~$\approx 0.65$, nearly independent of the nature of the WIMP.   The best fit still disfavors the absence of dark radiation (\Deln~= 0 at $\sim 95\%$ confidence), while allowing for the presence of a sterile neutrino (\Deln~= 1 at $\la 1\,\sigma$).  For all cases considered here, the lithium problem persists.  These results, presented at the Rencontres de l'Observatoire de Paris 2013 - ESO Workshop and summarized in these proceedings, are based on \citet{kngs}.
}
\maketitle{}

\section{Introduction}

Late in the early evolution of the Universe, after the \epm pairs have annihilated, the only remaining standard model (SM) particles are the CMB photons and the three relic neutrinos ($\nu_{e}$, $\nu_{\mu}$, $\nu_{\tau}$).  At these early epochs the Universe is ``radiation dominated", and after the \epm pairs have annihilated, the energy density may be written as $\rho_{\rm R} = \rho_{\gamma} + 3\,\rho_{\nu}$, where $3\,\rho_{\nu}$ accounts for the contributions from the three, SM neutrinos.  In addition to the SM neutrinos, there may be additional, beyond the standard model particles that, like the SM neutrinos, are extremely light ($\la 10\,{\rm eV}$) and very weakly interacting.  During the early (or, even, relatively late) evolution of the Universe these ``extra", neutrino-like particles, so called ``equivalent neutrinos", will contribute to the energy density, which controls the early Universe expansion rate.  If \Deln~counts the contribution of equivalent neutrinos, often referred to as ``dark radiation", $\rho_{\rm R} = \rho_{\gamma} + (3 + \Deln)\,\rho_{\nu}$.  The contribution to \Deln~of an equivalent neutrino that decouples along with the SM neutrinos (at $T = T_{\nu d}$) will be \Deln~= 1 for a Majorana fermion (\eg, a sterile neutrino), \Deln~= 2 for a Dirac fermion or, \Deln~= 4/7 for a real scalar.  In general, \Deln~is an integer (fermions) or an integer multiple of 4/7 (bosons).  However, an equivalent neutrino that is more weakly interacting than the SM neutrinos, will have decoupled earlier in the evolution of the Universe and its contribution to \Deln~will be suppressed by the heating of the SM neutrinos (and photons) when the heavier SM particles decay and/or annihilate.  Therefore, in principle, there is no reason that \Deln~should be an integer or an integer multiple of 4/7 (for further discussion see \citet{steig13}; for a specific example of three, very weakly coupled, right-handed neutrinos, see \citet{haim}).

After the SM neutrinos have decoupled, when $T = T_{\nu d} \approx 2 - 3\,{\rm MeV}$, the \epm pairs annihilate, heating the photons but not the already decoupled neutrinos.  Prior to neutrino decoupling (and \epm annihilation), the neutrinos, \epm pairs, and the photons are in equilibrium at the same temperature, $T_{\nu} = T_{e} = T_{\gamma}$ but, after \epm annihilation, the photons are hotter than the relic neutrinos.  In the simplest, textbook discussions, it is assumed that the neutrinos decoupled instantaneously and that the electrons were effectively massless at neutrino decoupling, when $T_{e} = T_{\nu d}$.  With these approximations, the late time (after \epm annihilation is complete) ratio of neutrino and photon temperatures is $(T_{\nu}/T_{\gamma})_{0} = (4/11)^{1/3}$ and the ratio of energy densities in one species of neutrino ($\rho^{0}_{\nu}$) and the photons is $(\rho^{0}_{\nu}/\rho_{\gamma})_{0} = 7/8\,(T_{\nu}/T_{\gamma})^{4}_{0} = 7/8\,(4/11)^{4/3}$.  However, at neutrino decoupling $m_{e}/T_{\nu d} \approx 0.2 \neq 0$ and, $\rho_{\nu}$ differs (by a small amount) from $\rho^{0}_{\nu}$ \citep{steig13}.  Furthermore, the neutrinos don't decouple instantaneously and, while the neutrinos are partially coupled they share some (a small amount) of the energy released by \epm annihilation \citep{mangano}.  These effects can be accounted for by introducing \neff, the ``effective number of neutrinos", where $\rho_{\rm R} \equiv \rho_{\gamma} + {\rm N}_{\rm eff}\,\rho^{0}_{\nu}\,,$ so that,
\beq
{\rm N}_{\rm eff} = 3\,\bigg[{11 \over 4}\bigg({T_{\nu} \over T_{\gamma}}\bigg)^{3}_{0}\bigg]^{4/3}\bigg(1 + {\Deln \over3}\bigg)\,.
\eeq
Assuming instantaneous neutrino decoupling and that $m_{e} \ll T_{\nu d}$, \neff~= 3 + \Deln.  Assuming instantaneous decoupling but correcting for the finite electron mass, \neff~$\approx 3.02(1 + \Deln/3)$ \citep{steig13}.  Accounting for non-instantaneous neutrino decoupling and for the finite electron mass, \neff~$\approx 3.05(1 + \Deln/3)$ \citep{mangano}.  In addition, in this case there is a very small, but not entirely negligible correction to the BBN predicted primordial helium abundance \citep{mangano}.

\begin{figure}[]
\resizebox{\hsize}{!}{\includegraphics[clip=true]{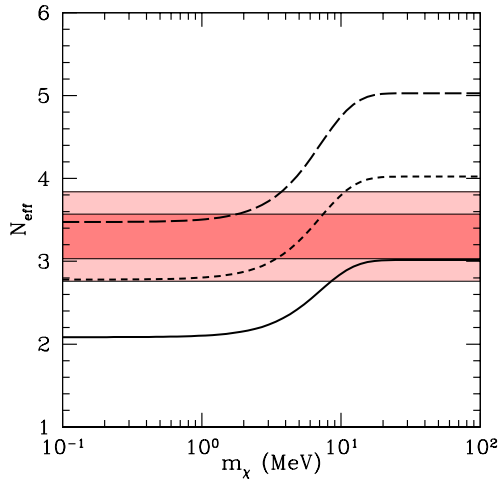}}
\caption{N$_{\rm eff}$ is shown as a function of the WIMP mass for \Deln~equivalent neutrinos, for the case of a Majorana fermion WIMP.  The solid  curve is for \Deln~= 0, the short dashed curve is for \Deln~= 1, and the long dashed curve is for \Deln~= 2.  The horizontal, red bands are the Planck CMB 68\% and 95\% allowed ranges.  This figure is from \citet{kngs}; an earlier version is in \citet{steig13}.}
\label{fig:neffvsmchi}
\end{figure}

So far, the possibility of a very light, weakly interacting, massive particle, a WIMP $\chi$, has been ignored.  The difference between a WIMP and an equivalent neutrino is that a WIMP remains thermally coupled to the SM particles after it has become non-relativistic and when it begins annihilating and, its annihilation heats the remaining SM particles (either the photons and, possibly, the \epm pairs if the WIMP couples electromagnetically or, the SM neutrinos if the WIMP only couples to them).  Note that in the analysis and discussion here, the WIMP {\bf need not} be the dark matter; it could be a sub-dominant component of the dark matter ($\Omega_{\chi} < \Omega_{\rm CDM}$).  Here we specialize to the case of a light WIMP coupled only to the photons and \epm pairs.  The relevant role played by such a light WIMP is that its annihilation heats the photons relative to the decoupled SM neutrinos, changing (reducing) $(T_{\nu}/T_{\gamma})_{0}$.  In this case, \neff~is a function of \mchi~(see \citet{steig13} and references therein).  The expansion rate of the early Universe, the Hubble parameter $H$, is controlled by the energy density ($H \propto \rho_{\rm R}^{1/2}$), so any modification of \neff~will be reflected in a non-standard expansion rate (\eg, during BBN).  Extremely light WIMPs (\mchi~$\la m_{e}$) will annihilate so late that, if their annihilation produces photons, they will modify the baryon-to-photon ratio ($\eta_{10} = 10^{10}(n_{\rm B}/n_{\gamma})_{0} = 273.9\,\omb$) during or after BBN.  BBN can probe \neff~as well as the universal ratio of baryons-to-photons.  At late times, \eg, at recombination, the CMB can also probe \omb~and \neff.  As independent probes of the effective number of neutrinos (\neff) or the number of equivalent neutrinos (\Deln) and the universal baryon density (\omb~or $\eta_{10}$), BBN and the CMB can help to break the degeneracies among these parameters and the WIMP mass (and spin/statistics) and to constrain their allowed ranges (see, \citet{steig13} and Fig.\,\ref{fig:neffvsmchi}).

\subsection{Planck CMB Constraints}

In the analysis in \citet{kngs}, whose results are described and summarized here, the CMB constraints on \omb~and \neff~are adopted from the Planck $\Lambda\mathrm{CDM}+N_\mathrm{eff}$ fit including BAO \citep{planck}.  The correlations between these quantities have been included in our analysis.  For our analysis we have adopted \omb~$= 0.0223 \pm 0.0003$ ($\eta_{10} = 6.11 \pm 0.08$) and \neff~$= 3.30 \pm 0.27$.  In Fig.\,\ref{fig:neffvsmchi}, the Planck 68\% and 95\% constraints on \neff~are shown as a function of the WIMP mass (the CMB constraints are independent of the WIMP mass).  Also shown are the curves corresponding to \neff~as a function of \mchi~for a Majorana fermion WIMP and for three choices of the number of equivalent neutrinos.  The behavior seen here is qualitatively similar for a Dirac or scalar WIMP (see, \eg, \citep{steig13}).  This figure illustrates the degeneracies between \neff~and \mchi.  For example, for \Deln~= 0 the CMB can set a {\bf lower} bound to \mchi.  In contrast, for \Deln~= 1\,(2), high values of \mchi~are excluded.

\subsection{BBN Constraints}

\begin{figure}[]
\resizebox{\hsize}{!}{\includegraphics[clip=true]{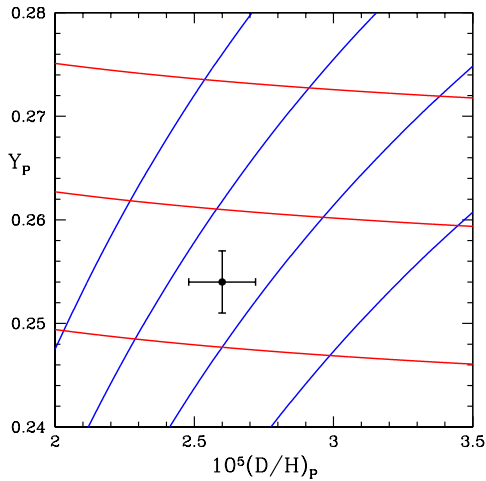}}
\caption{\footnotesize BBN predicted curves of constant baryon-to-photon ratio and equivalent number of neutrinos in the \Yp~-- $y_{\rm DP}$ plane.  From left to right (blue), $\eta_{10} = 7.0, 6.5, 6.0, 5.5$.  From bottom to top (red) \Deln~= 0, 1, 2.  Also shown by the filled circle and error bars are the observationally inferred values of \Yp~and $y_{\rm DP}$ adopted here (see the text).}
\label{fig:YvsD}
\end{figure}

Of the light nuclides produced during BBN, D and \4he are the relic nuclei of choice.  To account for, or minimize, the post-BBN contributions to the primordial abundances, observations at high redshift (z) and/or low metallicity (Z) are preferred.  Deuterium (and hydrogen) is observed in high-z, low-Z, QSO absorption line systems and helium is observed in relatively low-Z, extragalactic \hii~regions.  Even so, it may still be necessary to correct for any post-BBN nucleosynthesis that may have modified their primordial abundances.  The post-BBN evolution of D and \4he is simple and monotonic.  As gas is cycled through stars, D is destroyed and \4he produced.  Finally, D and \4he provide complementary probes of the parameters of interest.  $y_{\rm DP} \equiv 10^{5}{\rm (D/H)}_{\rm P}$ is mainly sensitive to the baryon density at BBN (\omb) and is less sensitive to \Deln.  In contrast, the \4he mass fraction, \Yp, is very insensitive to \omb, but is quite sensitive to \Deln.  This complementary, nearly orthogonal, dependence of D and \Yp~on $\eta_{10}$ and \Deln~is illustrated in Fig.\,\ref{fig:YvsD}.  For the analysis here (and in \citet{kngs}), we have adopted, $y_{\rm DP} = 2.60 \pm 0.12$ \citep{pettini} and \Yp~$= 0.254 \pm 0.003$ \citep{izotov}. 

In contrast, \3he has a more complicated, model dependent, post-BBN evolution and has only been observed in the relatively metal-rich interstellar medium of the Galaxy.  In addition, its BBN-predicted abundance is less sensitive to \omb~and \Deln~than that of D.  \3he is not used in our BBN analysis but, we have confirmed that its observationally inferred primordial abundance \citep{rood} is in good agreement with our BBN-predicted results.  \7li suffers from some of the same issues as \3he.  Its post-BBN evolution is complicated and model dependent.  Although, in principle, \7li could be as useful as D in constraining \omb~(and, to a lesser extent, \Deln), there is the well known ``lithium problem" (see, \eg, \citet{fields} and \citet{spite} for recent reviews) that, as will be seen below, persists.  In the BBN analyses, with and without a light WIMP, only D and \4he are used to constrain \omb~and \Deln~(or, \neff) and these BBN constraints are compared to the independent constraints from the CMB.

\begin{figure}[]
\resizebox{\hsize}{!}{\includegraphics[clip=true]{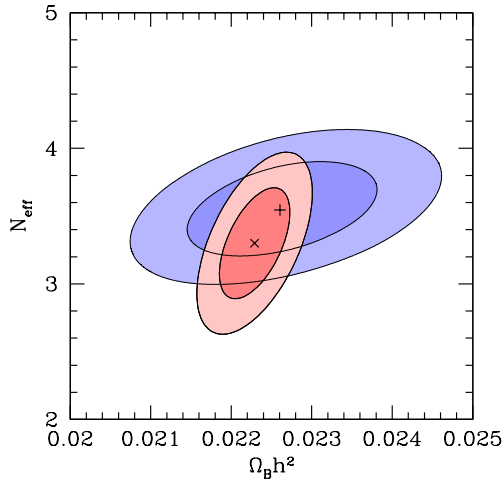}}
\caption{\footnotesize A comparison of the the 68\% (darker) and 95\% (lighter) contours in the \neff~--~\omb~plane derived separately from BBN (blue) and the CMB (pink).  The ``$\times$'' symbol marks the best fit CMB point and the ``$+$'' is the best fit BBN point.}
\label{fig:neffvsomb0}
\end{figure}

\section{BBN Without A Light WIMP}

\begin{figure}[]
\resizebox{\hsize}{!}{\includegraphics[clip=true]{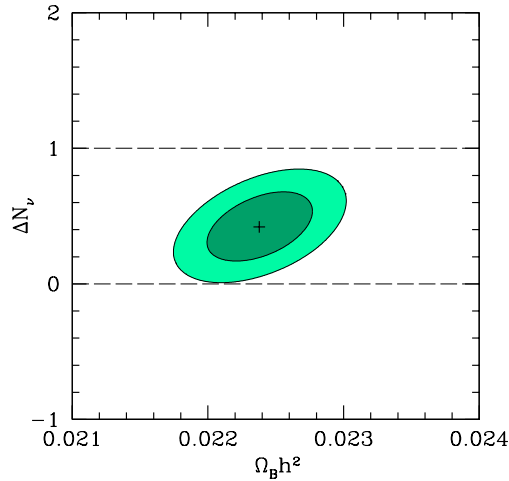}}
\caption{\footnotesize The  joint BBN+CMB 68\% (darker) and 95\% (lighter) contours in the number of equivalent neutrinos (\Deln) -- baryon density (\omb) plane.  The ``$+$'' symbol marks the best fit point.  SBBN corresponds to \Deln~= 0 and a sterile neutrino corresponds to \Deln~= 1, as shown by the dashed lines.}
\label{fig:delnvsomb1}
\end{figure}

In the absence of a light WIMP the BBN-predicted primordial abundances depend on only two parameters, the baryon-to-photon ratio ($\eta_{10}$ or, \omb) and the number of equivalent neutrinos (\Deln).  In the absence of a light WIMP the effective number of neutrinos and the number of equivalent neutrinos are related by \neff~$ = 3.05\,(1 + \Deln/3)$.  With two, independent relic abundances (D and \4he), BBN can constrain these two parameters.  This is illustrated in Fig.\,\ref{fig:YvsD}.  For the abundances adopted here, we find from BBN (without a light WIMP), $\eta_{10} = 6.19 \pm 0.21$ (\omb~$= 0.0226 \pm 0.0008$) and \Deln~$= 0.51 \pm 0.23$, corresponding to \neff~$= 3.56 \pm 0.23$ (accounting for round-off).  The BBN 68\% and 95\% contours in the \neff~-- \omb~plane, along with the best fit point, are shown in Fig.\,\ref{fig:neffvsomb0}, where they are compared to the corresponding contours (and best fit point) for these parameters inferred from the Planck CMB data \citep{planck}.  As Fig.\,\ref{fig:neffvsomb0} reveals, in the absence of a light WIMP, there is excellent agreement between BBN and the CMB.  This motivates (justifies) a joint BBN + CMB analysis, resulting in (for the joint fit) $\eta_{10} = 6.13 \pm 0.07$ (\omb~$= 0.0224 \pm 0.0003$) and \neff~$= 3.46 \pm 0.17$ (\Deln~$= 0.40 \pm 0.17$).  However, as may be seen from Fig.\,\ref{fig:delnvsomb1}, this joint BBN + CMB fit favors neither standard BBN (SBBN: \Deln~= 0), nor the presence of a sterile neutrino (\Deln~= 1).  SBBN is disfavored at $\sim 2.4\,\sigma$ and a sterile neutrino is disfavored at $\sim 3.5\,\sigma$.  

As for lithium, for the joint BBN + CMB parameter values the BBN predicted \7li abundance is ${\rm A(Li)} \equiv 12 + {\rm log\,(Li/H)} = 2.72 \pm 0.04$, to be compared with the observationally inferred ``Spite Plateau" abundance of ${\rm A(Li)} = 2.20 \pm 0.06$ \citep{spite}.  The lithium problem, the factor of $\sim 3$ difference between predictions and observations, persists.

\section{BBN With A Light WIMP}

Although BBN and the CMB are in excellent agreement in the absence of a light WIMP, we are interested in investigating the constraints they can set on the mass of such a WIMP and also, how its presence changes the parameter constraints discussed in the previous section.  The presence of a light WIMP can effect BBN (and the CMB) in several ways, provided it is sufficiently light.  For example, a very light WIMP might be mildly relativistic at BBN (or, prior to BBN, when the neutron-to-proton ratio is being set), contributing to the total energy density (similar to an equivalent neutrino) and speeding up the expansion rate.  A faster expansion generally increases the neutron-to-proton ratio at BBN, leading to the production of more \4he.  Also, such a very light WIMP might annihilate during or after BBN and the photons produced by its annihilation will change the baryon-to-photon ratio from its value during BBN.  The baryon-to-photon ratio at present may differ from its value at BBN affecting, mainly, the BBN D abundance.  The effects on the BBN light element yields in the presence of a light WIMP but, neglecting any equivalent neutrinos (\Deln~$\equiv 0$), were investigated by \citet{ktw} and \citet{serpico} and, more recently, by \citet{boehm}.  In \citet{kngs} those BBN calculations were extended to allow for the presence of dark radiation (\Deln~$\neq 0$).  In this case, there are three free parameters.  In addition to the baryon density ($\eta_{10}$ or \omb) and the number of equivalent neutrinos (\Deln), the light WIMP mass is allowed to vary, modifying the connection between \neff~and \Deln,
\beq
\neff~= {\rm N}_{\rm eff}^{0}(\mchi)(1 + \Deln/3)\,,
\eeq
and producing time-dependent effects on the weak rates and the expansion rate during BBN.  As already noted by \citet{ktw}, \citet{serpico} and, \citet{boehm}, for an electromagnetically coupled light WIMP, as \mchi~decreases below $\sim 20\,{\rm MeV}$, the BBN predicted D abundance decreases monotonically, while the \4he abundance first decreases (very slightly) and then increases monotonically.  For a more detailed discussion of the physics controlling this modified BBN, especially the non-monotonic behavior of \Yp~and its connection to the temperature dependence of the neutron -- proton interconversion reactions, see \citet{kngs}.

\begin{figure}[]
\resizebox{\hsize}{!}{\includegraphics[clip=true]{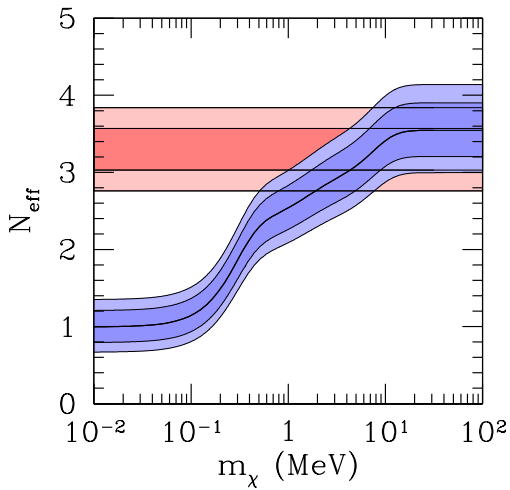}}
\caption{\footnotesize The CMB and BBN constraints on \neff~as a function of the light WIMP mass, \mchi.  The horizontal, pink bands show the 68\% and 95\% ranges from the Planck CMB results.  The blue bands show the corresponding BBN ranges.  The black curve through the middle of the blue bands shows the values of \neff~as a function of \mchi~for which the BBN predicted D and \4he abundances agree exactly with the observationally inferred abundances adopted here.}
\label{fig:neffvsmchi1}
\end{figure}

\begin{figure}[]
\resizebox{\hsize}{!}{\includegraphics[clip=true]{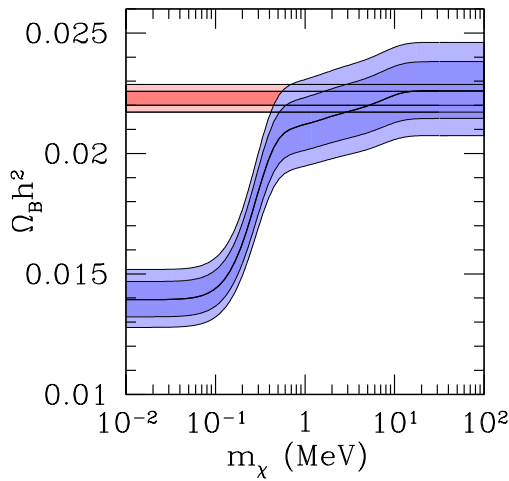}}
\caption{\footnotesize As in Fig.\,\ref{fig:neffvsmchi1}, the CMB and BBN constraints on \omb~as a function of the light WIMP mass, \mchi.}
\label{fig:ombvsmchi1}
\end{figure}

With three parameters and two observables ($y_{\rm DP}$ and \Yp), BBN is underconstrained.  For each choice of \mchi, a pair of $\eta_{10}$ and \Deln~parameters can be found so that BBN predicts -- exactly -- the observed D and \4he abundances.  This is illustrated in Figs.\,\ref{fig:neffvsmchi1} and \ref{fig:ombvsmchi1}, which show \neff~and \omb~as functions of the WIMP mass, as inferred from the CMB (\neff~and \omb~are independent of \mchi) and from BBN.  These figures show how the degeneracy illustrated in Fig.\,\ref{fig:neffvsmchi} can be broken by combining constraints from the CMB with those from BBN.  

\begin{figure}[]
\resizebox{\hsize}{!}{\includegraphics[clip=true]{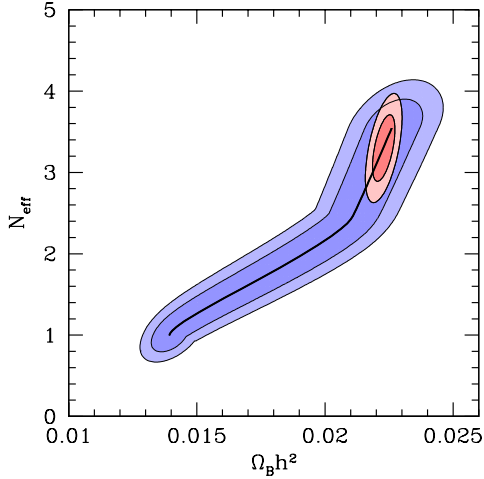}}
\caption{\footnotesize The CMB and BBN constraints on \neff~as a function of the the baryon density \omb~(combining the results shown in Figs.\,\ref{fig:neffvsmchi1} \& \ref{fig:ombvsmchi1}).}
\label{fig:neffvsomb}
\end{figure}

A comparison of the BBN and CMB constraints on \neff~and \omb~is shown in Fig.\,\ref{fig:neffvsomb}.  The independent and complementary BBN and CMB results are in excellent agreement, over the range in \neff~and \omb~defined by the Planck CMB constraints.  As a result, the BBN and CMB results may be combined in a joint analysis to identify the allowed 68\% and 95\% ranges in the \neff~(or, \Deln) -- \omb~plane.  This joint analysis \citep{kngs} finds \neff~$= 3.30 \pm 0.26$ and \omb~$= 0.0223 \pm 0.0003$ ($\eta_{10} = 6.11 \pm 0.08$), consistent with the CMB results alone, but with slightly smaller uncertainties.  The new results from this joint analysis for \Deln~as a function of \omb~are shown in Fig.\,\ref{fig:delnvsomb}.  For the joint fit, \Deln~$= 0.65^{+0.46}_{-0.35}$.  Note that these figures and the numerical results cited here are for the case of a Majorana fermion WIMP.  Very similar results are found for a Dirac fermion or for a real or complex scalar WIMP (see Table 1 in \citet{kngs}).

Allowing for a light WIMP, the joint CMB + BBN comparison excludes light WIMPs with masses $\la 0.5 - 5\,{\rm MeV}$.  The best joint fit WIMP mass is found to be \mchi~$\approx 5 - 10\,{\rm MeV}$, depending on the nature of the WIMP.  However, very nearly independently of the nature of the WIMP, the best fit for the dark radiation is \Deln~$\approx 0.65$ in all cases (see Fig.\,10 and Table 1 in \citet{kngs}).  While \Deln~= 0 is still disfavored at $\sim 95\%$ confidence, in the presence of light WIMP, a sterile neutrino (but, not two sterile neutrinos!) is now permitted.  Since the no light WIMP case is a good fit to the BBN and CMB data, there is no upper bound to the WIMP mass.

It is interesting that for the WIMP masses allowed by the joint BBN + CMB fit (including the high WIMP mass limit -- the no light WIMP case), the BBN predicted lithium abundance lies in the range ${\rm A(Li)} = 2.72 \pm 0.04$ (see Fig.\,13 in \citet{kngs}), still a factor of $\sim 3$ larger than the observationally inferred Spite Plateau value of A(Li) $= 2.20 \pm 0.06$ \citep{spite}.  A light WIMP does not help to alleviate (indeed, it reinforces) the lithium problem.

\section{Summary And Conclusions}

\begin{figure}[]
\resizebox{\hsize}{!}{\includegraphics[clip=true]{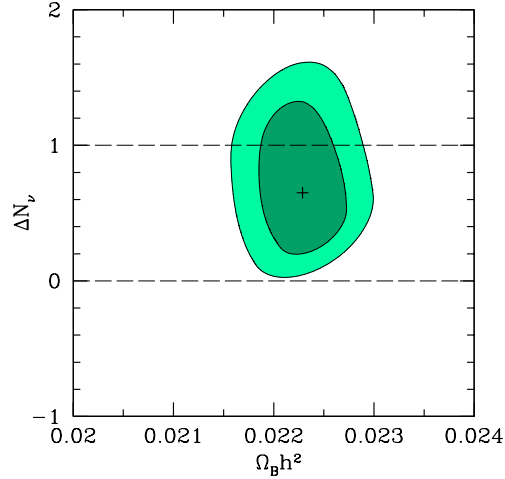}}
\caption{\footnotesize The joint CMB + BBN constraints on dark radiation (\Deln) as a function of the baryon density (\omb).  The dashed lines indicate the absence of dark radiation (\Deln~= 0) and the presence of a sterile neutrino (\Deln~= 1).}
\label{fig:delnvsomb}
\end{figure}

In the absence of a light WIMP the effective number of neutrinos and the number of equivalent neutrinos are simply related, \neff~$= 3.05(1 + \Deln/3)$ and the Planck CMB data alone, \neff~$= 3.30 \pm 0.27$ \citep{planck}, constrains \Deln~$= 0.25 \pm 0.27$, consistent with the absence of dark radiation at $\la 1\,\sigma$ (and, inconsistent with a sterile neutrino at $\sim 2.8\,\sigma$).  The CMB alone also provides a constraint on the universal baryon density, \omb~$= 0.0223 \pm 0.0003$ ($\eta_{10} = 6.11 \pm 0.08$).  For \Deln~= 0 and the Planck value of the baryon density, BBN (SBBN) predicts the primordial D abundance to be $y_{\rm DP} = 2.48 \pm 0.05$, in excellent agreement with the observationally inferred value of $y_{\rm DP} = 2.60 \pm 0.12$ \citep{pettini}.  However, for this combination of \omb~and \Deln, the SBBN predicted primordial helium abundance is \Yp~$= 0.2472 \pm 0.0005$, which is $\sim 2.3\,\sigma$ away from the observationally inferred value, \Yp~$= 0.254 \pm 0.003$ \citep{izotov}.  Independent of the CMB, in the absence of a light WIMP BBN provides independent constraints on \Deln~(\neff) and \omb.  BBN alone finds \Deln~$= 0.51 \pm 0.23$ (\neff~$= 3.56 \pm 0.23$) and \omb~$= 0.0226 \pm 0.0008$ ($\eta_{10} = 6.19 \pm 0.21$).  Within the errors, the BBN and CMB constraints on \neff~and \omb~(\Deln~and $\eta_{10}$) are in excellent agreement.  However, neither SBBN (\Deln~= 0) nor a sterile neutrino (\Deln~= 1) is favored by the combined BBN + CMB analysis, which finds \Deln~$= 0.40 \pm 0.17$.  Indeed, in the absence of a light WIMP, \Deln~= 0 is disfavored at $\sim 2.4\,\sigma$ and \Deln~= 1 is disfavored at $\sim 3.5\,\sigma$. The joint BBN + CMB analysis predicts a primordial lithium abundance A(Li)~$= 2.72 \pm 0.04$, a factor of $\sim 3$ higher than the observationally inferred, primordial value; the ``lithium problem" persists.

In the presence of a sufficiently light WIMP (\mchi~$\la 20\,{\rm MeV}$) the CMB results are unchanged (although the connection between \neff~and \Deln~is modified depending on the WIMP mass).  In this case \neff~$= {\rm N}^{0}_{\rm eff}(\mchi)(1 + \Deln/3)$ and there is a degeneracy between the CMB constraints on \neff~and \Deln~(and \mchi).  As may be seen from Fig.\,\ref{fig:neffvsmchi}, for some choices of \Deln, the CMB constraint on \neff~sets a lower limit to \mchi, while for other choices the CMB sets an upper limit to the WIMP mass.  The independent constraints from BBN can help to break these degeneracies.  In the presence of a light WIMP BBN depends on three parameters: \Deln, \neff, \omb~(or, \mchi, \Deln, $\eta_{10}$) but, there are only two BBN constraints from D and \4he.  For each choice of \mchi, there is always a pair of \Deln~and $\eta_{10}$ values for which BBN predicts -- exactly -- the observationally inferred primordial D and \4he abundances.  However, the corresponding BBN inferred values (and ranges) of \neff~and \omb~need not agree with the values (and ranges) set by the CMB.  By comparing the BBN and CMB constraints, the degeneracies may be broken.  In this way a lower bound, as well as a best fit value, of the WIMP mass is found (depending on the nature of the WIMP).  For the case of a Majorana fermion WIMP shown in the figures, \mchi~$\ga 1.7\,{\rm MeV}$ and the best fit is for a WIMP mass \mchi~$ = 7.9\,{\rm MeV}$.  Depending on the nature of the WIMP, the lower bound to \mchi~ranges from $\sim m_{e}$ to $\sim 10\,m_{e}$, while the best fit WIMP masses lie in the range $\sim 5 - 10\,{\rm MeV}$ (see \citet{kngs}).  In all cases, very nearly independent of the nature of the WIMP, \Deln~$\approx 0.65$.  While the joint BBN + CMB analysis finds essentially the CMB values for \neff~and \omb, the presence of an additional, free parameter relaxes the constraints (increases the error) on \Deln~(compared to the no light WIMP case).  Now, a sterile neutrino is permitted at $\la 68\%$ confidence (see Fig.\,\ref{fig:delnvsomb}).  However, the absence of dark radiation (\Deln~= 0) is still disfavored at $\sim 95\%$ confidence.  For the joint BBN + CMB analysis the BBN predicted primordial lithium abundance is A(Li)~$= 2.73 \pm 0.04$, a factor of $\sim 3$ higher than the observationally inferred value.  Even in the presence of a light WIMP, the lithium problem persists.

It should be noted that at this meeting R.\ Cooke presented new results on the primordial abundance of deuterium, $y_{\rm DP} = 2.53 \pm 0.04$ \citep{cooke}.  Although the new central value agrees very well with the earlier, \citet{pettini} result adopted here, the new uncertainty is smaller by a factor of three.  In the analysis described here (and, in more detail in \citet{kngs}), this small change in the primordial deuterium abundance has the effect of increasing $\eta_{10}$ by $\sim 0.1$ and decreasing \Deln~by $\sim 0.01$.  These small changes, well within the errors, leave the results and conclusions presented here unaffected.

\begin{acknowledgements}
We are grateful to the Ohio State University Center for Cosmology and Astro-Particle Physics for hosting K.\,M.\,N's visit during which most of the work described here was done.  K.\,M.\,N is pleased to acknowledge support from the Institute for Nuclear and Particle Physics at Ohio University.  G.\,S.~is grateful for the hospitality provided by the Departamento de Astronomia of the Instituto Astron$\hat{\rm o}$mico e Geof\' \i sico of the Universidade de S\~ao~Paulo, where these proceedings were written.  The research of G.\,S.~is supported at OSU by the U.S.~DOE grant DE-FG02-91ER40690.  

\end{acknowledgements}

\bibliographystyle{aa}

\end{document}